\documentclass[12pt]{article}
\usepackage{sbc-template}

\usepackage{graphicx,url} 

\usepackage[utf8]{inputenc}
\usepackage{hyperref}

\title{Towards Multi-Criteria Prioritization of Best Practices in Research Artifact Sharing
\footnote{Open peer review artifacts for this paper are available at \url{https://zenodo.org/communities/opensciense2021}}
}

\author{Carlos Diego Nascimento Damasceno\inst{1}, Isotilia Costa Melo\inst{2,3}, Daniel Strüber\inst{1}}

\address{Faculty of Science (iCIS) - Radboud University Nijmegen, the Netherlands
\nextinstitute
  Sao Carlos Engineering School (EESC) - University of Sao Paulo (USP), Brazil
\nextinstitute
  Universidad Adolfo Ibáñez (UAI), Chile
  \email{\{d.damasceno,d.strueber\}@cs.ru.nl, isotilia@gmail.com}
}

\begin{document} 

\maketitle

\begin{abstract}
Research artifact sharing is known to strengthen the transparency of scientific studies. However, in the lack of common discipline-specific guidelines for artifacts evaluation, subjective and conflicting expectations may happen and threaten artifact quality. In this paper, we discuss our preliminary ideas for a framework based on quality management principles (5W2H) that can aid in the establishment of common guidelines for artifact evaluation and sharing. Also, using the Analytic Hierarchy Process, we discuss how research communities could join efforts to aid the guidelines' adequacy to research priorities. These combined methodologies constitute a novelty for software engineering research which can foster research software sustainability. 
\end{abstract}

\section{Introduction}

Research artifacts are known to allow others to build ideas upon existing knowledge, adopt novel ideas in practice, and increase the likelihood of citations.
By "artifact" we mean any digital object created to be used as part of a study or generated in an experiment \cite{acm_artifact_2020}. This definition covers software systems, scripts used to run experiments, input datasets, raw data collected in the experiment, or scripts used to analyze results.

While artifact sharing is known to be a social good \cite{timperley_understanding_2020}, the creation and maintenance of high-quality research artifacts can be challenging as researchers may have different expectations toward \textit{artifact quality} \cite{van_eeuwijk_research_2021}.
Also, the current lack of discipline-specific guidelines for research data management (RDM) \cite{marjan_grootveld_openaire_2018} opens the opportunity to misunderstandings on the potential of artifacts and threats to the sustainability of research projects \cite{hermann_community_2020}.
To address these issues, we envision the need for studies experimenting with project management principles for artifact quality management in software engineering (SE) research. 

In this paper, we present initial thoughts on a generic framework for SE research artifact quality management.
Our framework is envisioned as a guide for artifact stakeholders (e.g., users, authors, reviewers, artifact evaluation communities - AECs) to identify essential concerns (i.e., research practices and frequently asked questions) in artifact quality management and prioritize them based on project constraints.
Particularly, we aim to extend the work of \cite{damascenostrueber2021_qa_mde} beyond its initial scope.

We envision that their work \cite{damascenostrueber2021_qa_mde} can be improved with prioritization principles, 
such as the Analytic Hierarchy Process (AHP) \cite{saaty_models_2012}.
The AHP can provide a participatory model that accommodates the interests of the research community and AEC members in a mutually prioritized and agreed-upon guideline. 
Thus, we can encompass the research community and AEC members' viewpoints of what concerns should be given higher priority in artifact creation and review. 
Finally, we believe that the work of \cite{damascenostrueber2021_qa_mde} has the potential to help in other types of research artifacts and domains, e.g., datasets for machine learning \cite{Lindauer_practices_rnn2020}, data collected in surveys \cite{hui_reporting_2019}.

\section{Proposed Framework} \label{sec:aqm_se}

According to the Project Management Institute (PMI), 
project management is defined as the application of knowledge, skills, tools, and techniques to project activities to meet the project requirements \cite{pmi_guide_2017}. 
Among the project management knowledge areas, quality management is fundamental as it applies to all projects, regardless of the nature of their deliverables. 
The objectives of project quality management are to incorporate the organization’s quality policy regarding planning, managing, and controlling project and product quality requirements so that stakeholders’ expectations are met. 
In SE research, artifact stakeholders may include funding agencies, research collaborators, artifact users, or AEC members.

\begin{figure*}[ht!]
    \centering
    \includegraphics[width=\linewidth]{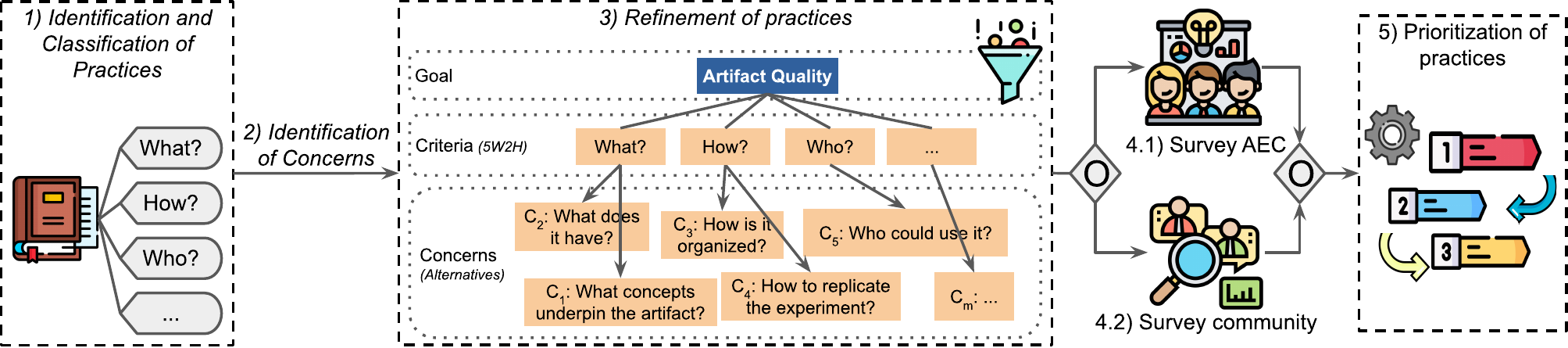}
    \caption{Proposed Framework for Design and Prioritization of Quality Guidelines}
    \label{fig:framework}
\end{figure*}

Artifact quality requirements are specific to the type of artifact being produced and the research domain. Thus, they should be identified and documented \textit{a priori}. 
The plan quality management is the process of identifying quality requirements and/or standards for a project and its artifacts and documenting how a project shall demonstrate compliance with such quality requirements and/or standards. Otherwise, it may have serious negative consequences for the project stakeholders.
With this in mind, in the next sections, we introduce a framework for designing and prioritizing quality guidelines for artifact sharing.
In Figure \ref{fig:framework}, we illustrate a schema of our proposed framework.

\subsection{Identification of Research Practices and Artifact Quality Concerns}

To understand the concept of artifact quality requirements, there is an extensive toolbox of methods \cite{tague_quality_2005} that can be used. 
Among them, the 5W2H (an acronym for \textbf{W}hat, \textbf{W}here, \textbf{W}hy, \textbf{W}ho, \textbf{W}hen, \textbf{H}ow, and \textbf{H}ow Much) constitutes a simple but powerful method for inquiring questions about a problem, as well as for research planning, analysis, and reviews. 

We see the 5W2H as a useful method to identify and document factual questions that artifact stakeholders may frequently inquire about an artifact and indicate \textit{What concerns should artifact stakeholders keep in mind?}. 
For example, some factual questions that could be asked are: 
\textit{\textbf{W}hat is the context of the development of this artifact?}
\textit{\textbf{W}here is this artifact hosted?}
\textit{\textbf{W}hy was this artifact created?}
\textit{\textbf{W}ho are the artifact authors?}
\textit{\textbf{W}hen did changes in this artifact happen?}
\textit{\textbf{H}ow to reproduce the experiment?}
\textit{\textbf{H}ow much RAM does it require?}.
To address these 5W2H questions, we assume there is a list of research best practices explicitly documented or implicitly known by a given research community. 
This list of practices could be cataloged in collaboration with domain experts, e.g., via community surveys, one-to-one interviews, or by means of literature review.

\subsection{Community Survey and Prioritization of Practices}

Once a set of useful factual questions is agreed upon and documented, 
time and resource constraints may limit the investments in creating and maintaining an artifact.
Thus, prioritization techniques could be helpful to rank artifact quality requirements, as done in traditional software requirements management \cite{pitangueiraetal2013_requirementsprioritiztion}. 

In the prioritization literature, the Analytic Hierarchy Process (AHP) stands out as a practical tool that can incorporate users' preferences for decision making through the pair-wise judgment of possible solutions for a given problem \cite{yooetal2009_testcase_ahp}. 
Particularly, the AHP constitutes an interesting model where artifact stakeholders can join efforts to categorize different quality concerns (i.e., factual questions) and research practices according to their relative importance towards the goal of sharing high-quality artifacts. 

Developed in the 1970s, the AHP is a structured multi-criteria decision-making approach that is underpinned by psychology and mathematics principles. 
In the AHP, individual domain-specific experts (e.g., AEC members, SE research community) estimate the relative importance of factors through pair-wise comparison (i.e., how much important is it to answer a given factual question?). 
Then, this prioritization criteria can be constrained to attend the preferences of reviewers or a research community by using mathematical consistency assumptions \cite{saaty_models_2012,colin_pesquisa_2000}.

\begin{figure*}[ht!]
    \centering
    \includegraphics[width=\linewidth]{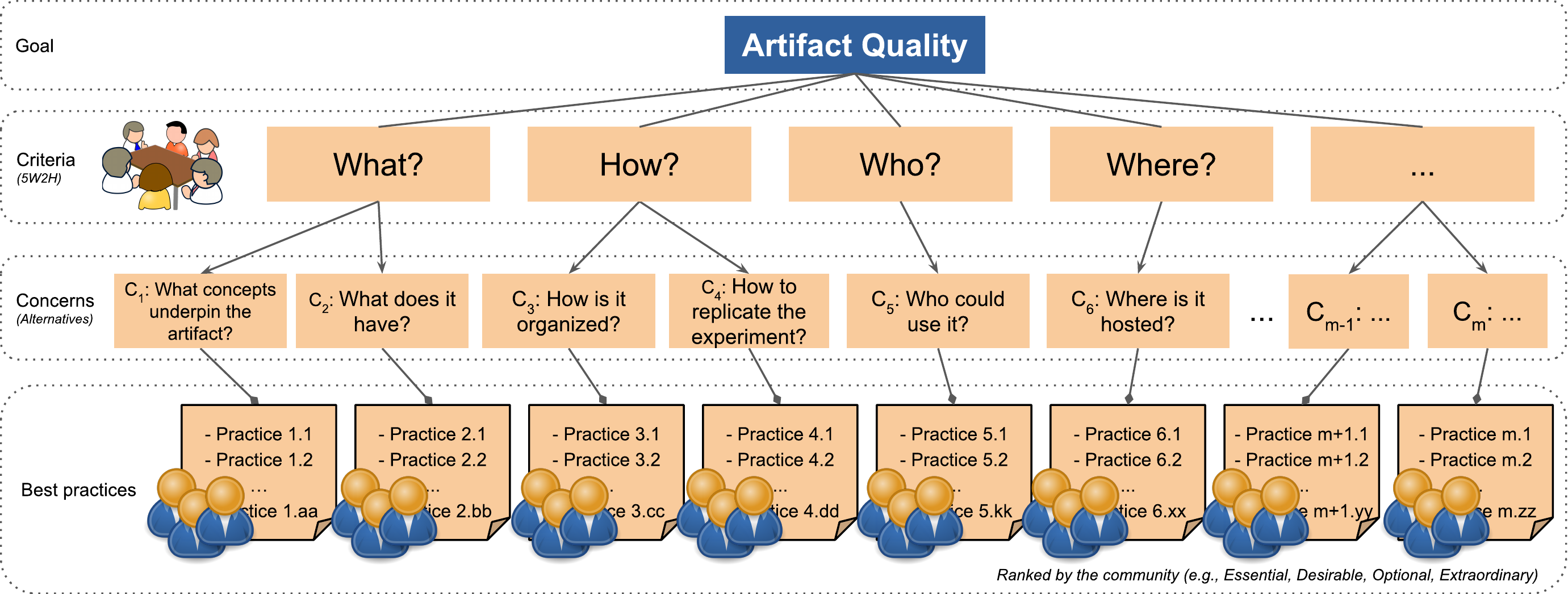}
    \caption{The AHP-Based Prioritization of Quality Guidelines}
    \label{fig:ahp_artifact}
\end{figure*}

In our framework, we envision the AHP as a practical means for ranking
\textit{Artifact quality concerns} and \textit{Research best practices}.
In Figure \ref{fig:ahp_artifact}, 
we provide a schema of our AHP-based solution to prioritizing artifact quality requirements and research practices. 

In our framework, we propose the AHP as a tool to find out “What is the hierarchical importance priority of the artifact quality concerns established through 5W2H method?”. 
These quality concerns should be agreed upon among AEC members as being \textbf{useful questions that artifact authors should provide answers to}. 
Once this set of quality concerns is agreed upon, AEC members shall individually indicate the relative importance of those questions through pair-wise comparisons. 
As recommended by the AHP method, experts should assign grades from one to nine to say which alternative out of a pair they consider as more important \cite{saaty_models_2012,colin_pesquisa_2000}. 
 
Second, the AHP shall be applied again to identify “What is the hierarchical importance priority of a list of best practices for artifact sharing?”. 
This should be answered by SE community members, also through pair-wise comparisons of research best practices. 
As in the ACM SIGSOFT Empirical Standards \cite{ralph_acm_2020}, community members can be asked to rank best practices according to their relative priority levels, 
e.g., Essential, Desirable, Extraordinary.

One advantage of segmenting this hierarchy is that AEC members can independently establish domain-specific or SE-wide artifact quality concerns, 
while the SE community can establish their own best practices used to address a given quality concern. 
On the other hand, considerations on the extenuation of experts during extensive pair-wise analysis should also be raised. 
To surpass this issue, AHP variants that rely on a reduced number of comparisons \cite{leal_ahpexpress_2020}, inter-cluster prioritization \cite{yooetal2009_testcase_ahp}, or metaheuristics \cite{bose_using_2020} could be explored as alternative prioritization methods.

In summary, using this framework, AEC members could tackle the need for high-quality research artifacts by bringing into the spotlight the most important 
\textit{research best practices for artifact sharing}
and
\textit{factual questions that researchers should concern about an artifact}.
The AHP is possibly an adequate approach once it promotes a hierarchical representation of the problem, considering different users’ judgments and preferences through the mathematical aggregation of priorities; and fosters the community engagement on reaching an agreement of what \textit{research practices} and \textit{artifact quality concerns} are of utmost importance for research software sustainability \cite{van_eeuwijk_research_2021}.

\section{Preliminary Results}

In the SE literature, there is a lack of domain-specific quality guidelines for artifact sharing. To the best of our knowledge,  \cite{damascenostrueber2021_qa_mde} have been the first to investigate quality guidelines for Model-Driven Engineering (MDE) research artifacts.

In MDE research, there is a limited number of data sets of benchmark models of diverse modeling languages and application domains \cite{basciani2015model}.
Moreover, the need for consolidated artifact sharing practices in MDE research has recently become more pronounced, as the community targets a broader use of artificial intelligence (AI) techniques.
To benefit from the advances in AI and deep learning, MDE researchers need to have access to larger open data sets and high confidence measures for their quality.
Thus, having more systematic artifact sharing practices would help to increase the impact of MDE research and support the thorough evaluation of MDE research tools and techniques.

To contribute towards the quality of MDE research artifacts, \cite{damascenostrueber2021_qa_mde} introduced a set of guidelines for artifact sharing specifically tailored to MDE research. 
To design these guidelines, they systematically analyzed general-purpose research practices for artifact sharing used by major SE venues, categorized them according to the 5W2H method, and tailored them to MDE research artifacts. 
Subsequently, the authors conducted an online survey among 90 researchers and practitioners with expertise in MDE. 
In this survey, participants were asked to rank each item of the proposed guideline as essential, desirable, or unnecessary; and evaluate them with respect to clarity, completeness, and relevance. 
In each of these dimensions, the proposed guidelines were assessed positively by more than 92\% of the participants. 

We believe that the results by \cite{damascenostrueber2021_qa_mde} can be extended by considering the multi-criteria prioritization of their guidelines. 
Prioritization techniques, as the AHP, may be helpful to manage artifact quality based on the relevance of both practices and quality concerns. 
Besides, we believe that their proposed framework could also be applied to other research domains, not exclusively MDE projects, and investigate how it could support artifact quality management in artifact evaluation.

\section{Final Remarks}

In this paper, we bring into the spotlight the limited knowledge on SE research artifact quality management.
As we discuss, artifact quality management has an impact on the sustainability of research artifacts. 
Hence, research artifact quality concerns and practices for artifact sharing should receive more attention from the SE research community. 
To fill this gap, we propose the combined application of quality management principles, the 5W2H method, and the AHP for designing and prioritizing artifact quality requirements.

For future works, we recommend a community-wide initiative towards identifying useful research practices and factual questions about artifact quality concerns, applying this proposed framework.
Once the community develops such a catalog of quality concerns and practices, AEC members could move towards a ranking of agreed-upon quality concerns and adopt the proposed guidelines in their daily routine and yearly conferences. 
This combined methodology constitutes a novelty for SE research project management that can foster more open and sustainable SE research and artifact development. 

\bibliographystyle{sbc}
\bibliography{sbc-template}

\end{document}